\documentclass[review]{elsarticle}

\usepackage{graphicx}
\usepackage{textcomp}
\usepackage{adjustbox}
\usepackage{xcolor}
\usepackage{booktabs}
\usepackage{multirow}
\usepackage[inkscapelatex=false]{svg}

\usepackage{amssymb}

\usepackage{amsmath}   
\usepackage{cleveref}  

\journal{}

\biboptions{authoryear}

\begin{document}

\begin{frontmatter}

\title{LLM-Enhanced Multimodal Fusion for Cross-Domain Sequential
Recommendation}
		
\author[firstaddress,secondaddress]{Wangyu Wu}\ead{wangyu.wu@liverpool.ac.uk}
\author[thirdaddress]{Zhenhong Chen}\ead{zcheh@microsoft.com}
\author[FourAddress]{Wenqiao Zhang}\ead{wenqiaozhang@zju.edu.cn}
\author[firstaddress,secondaddress]{Siqi Song}\ead{Siqi.Song22@student.xjtlu.edu.cn}
\author[firstaddress,secondaddress]{Xianglin Qiu}\ead{Xianglin.Qiu20@student.xjtlu.edu.cn}
\author[secondaddress]{Xiaowei Huang}\ead{xiaowei.huang@liverpool.ac.uk}
\author[firstaddress]{Fei Ma\corref{mycorrespondingauthor}}\ead{fei.ma@xjtlu.edu.cn}
\author[firstaddress]{Jimin Xiao\corref{mycorrespondingauthor}}
\cortext[mycorrespondingauthor]{Corresponding authors} \ead{jimin.xiao@xjtlu.edu.cn}

\address[firstaddress]{Xi'an Jiaotong-Liverpool University, Suzhou, China}
\address[secondaddress]{University of Liverpool, Liverpool, UK}
\address[thirdaddress]{Microsoft, Redmond, USA}
\address[FourAddress]{Zhejiang University, Hangzhou, China}

\begin{abstract}
Cross-Domain Sequential Recommendation (CDSR) predicts user behavior by leveraging historical interactions across multiple domains, capturing both intra- and inter-sequence item relationships. To further enhance the value of visual and textual data, we propose LLM-EMF, an innovative approach that incorporates Large Language Models(LLM) to enrich textual data and boosts recommendation performance by merging visual and textual information. Additionally, a multi-attention mechanism is designed to jointly learn single-domain and cross-domain preferences, effectively capturing complex user interests. Evaluations on four e-commerce datasets demonstrate that LLM-EMF outperforms existing methods in modeling cross-domain user preferences, highlighting the advantages of multimodal integration in sequential recommendation systems.

\end{abstract}

\begin{keyword}
Large Language Models, CDSR, CLIP-based Embeddings
\end{keyword}

\end{frontmatter}

\section{Introduction}
\label{sec:intro}
\begin{figure}[t]
\centering
\includegraphics[width=1.0\linewidth]{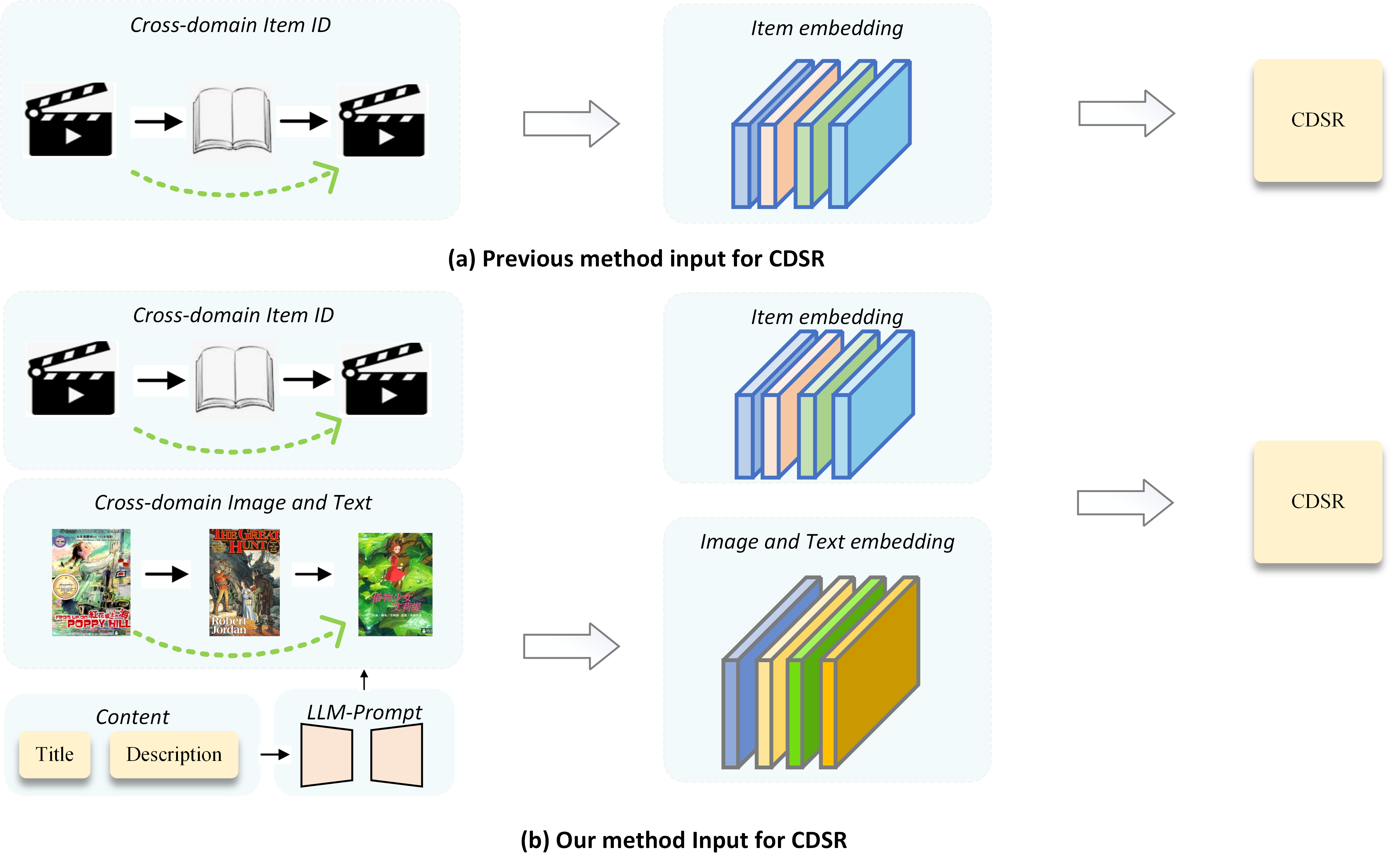}
\caption{(a) Traditional CDSR uses only item ID features. (b) Our LLM-EMF incorporates image and title information, enriching item representations.}
\label{fig:idea}
\end{figure}

Sequential Recommendation (SR) has become a widely adopted approach for modeling dynamic user preferences by analyzing historical interaction sequences~\cite{markovsr,transmarkov,markov,narm}. 
While SR effectively predicts the next item within a single domain, it suffers from data sparsity and domain bias, where limited interaction data and overfitting to domain-specific patterns hinder generalization. 
Cross-Domain Sequential Recommendation (CDSR)~\cite{pinet,Zhuangict,dagcn,kddsemi,cao2022contrastive,recguru,jiang2023structure} addresses these challenges by leveraging user behavior across multiple domains, enabling knowledge transfer and a more holistic modeling of user interests. 
However, existing CDSR methods still face three key limitations: (\emph{i}) they primarily focus on intra-domain dependencies, underexploring inter-domain relationships; (\emph{ii}) they insufficiently utilize multimodal information such as images and text; and (\emph{iii}) even LLM-based approaches~\cite{wei2024llmrec} rarely consider domain imbalance in cross-domain preference modeling. The difference between traditional CDSR and our approach is shown in Fig.~\ref{fig:idea}.

To address these gaps, we propose \textbf{LLM-Enhanced Multimodal Fusion for Cross-Domain Sequential Recommendation (LLM-EMF)}, a novel framework that integrates \emph{prompt-designed} Large Language Model (LLM) augmentation, multimodal fusion, and domain-balanced hierarchical attention. 
Unlike LLMRec~\cite{wei2024llmrec}, which augments textual features without explicitly mitigating domain imbalance, LLM-EMF generates domain-agnostic semantic attributes via LLM prompts to enhance cross-domain alignment. 
Compared with IFCDSR~\cite{wu2025image}, which fuses image and ID embeddings but lacks explicit textual enrichment, our framework unifies visual, textual, and ID-based embeddings. 
In contrast to MAN~\cite{lin2024mixed}, which applies mixed attention without balancing domain contributions, our hierarchical attention mechanism explicitly regulates domain influence, preventing high-frequency domains from dominating recommendations. 
To the best of our knowledge, LLM-EMF is the first CDSR framework that systematically integrates LLM-generated textual knowledge with multimodal fusion and explicit domain balance.

\textbf{Our key contributions are as follows:}
\begin{itemize}
    \item We design a \emph{prompt-based LLM augmentation} strategy to generate additional domain-agnostic textual attributes, improving semantic alignment between domains.
    \item We propose a \emph{multimodal fusion framework} that jointly models visual, textual, and ID embeddings via a hierarchical attention mechanism, capturing both intra- and inter-domain preferences while mitigating domain imbalance.
    \item We conduct extensive experiments on four reorganized e-commerce datasets, demonstrating that LLM-EMF consistently outperforms recent state-of-the-art baselines~\cite{wu2025image,lin2024mixed} across all evaluation metrics.
\end{itemize}

\section{Related Work}

\subsection{Sequential Recommendation}
Sequential recommendation~\cite{cai2024relation,SRs,liang2025graphical,wu2025impact} models user behavior as a sequence of time-sensitive items, aiming to predict the next item based on historical interactions. Early approaches, such as FPMC~\cite{rendle2010factorizing}, used first-order Markov chains to model temporal dependencies. With growing behavioral complexity, deep learning models—RNNs~\cite{GRU, LSTM}, CNNs~\cite{CNN}, and attention mechanisms~\cite{vaswani2017attention}—were integrated into recommender systems~\cite{gru4rec, DIEN, Caser, kang2018self, DIN}, capturing long-term dependencies and dynamic preferences. Methods like SURGE~\cite{SURGE} further reduced sequence complexity via metric learning for efficient large-scale retrieval. More recently, works such as DFAR~\cite{lin2023dual} and DCN~\cite{lin2022dual} have sought to capture deeper and more intricate relationships within sequential recommendation data using dual attention mechanisms and complex interaction modeling.

Our work extends these techniques with cross-domain learning, enabling knowledge transfer between domains and LLMs to leverage complementary signals, thereby improving generalization and adaptability.

\subsection{Cross-Domain Sequential Recommendation}
Unlike single-domain sequential recommendation, which models behavior within one domain, cross-domain recommender systems~\cite{xiao2025cross,CDR} address sparse data and cold-start issues by leveraging user behavior from multiple domains. Early methods~\cite{singh_relational_2008} assumed auxiliary behaviors from other domains could enhance the target domain. Transfer learning~\cite{Transfer} approaches (e.g., MiNet~\cite{MiNet}, CoNet~\cite{CoNet}, itemCST~\cite{itemCST}) transfer learned representations to boost target-domain recommendations.

Industrial settings often require improving performance in all domains simultaneously, motivating dual learning~\cite{long_dual_2012, he_dual_2016,hou2024invdiff,li2025bridge,li2024towards2,li2024distinct,li2024towards,liu2024fedbcgd,liuimproving}, as in DTCDR and DDTCDR, to enable mutual enhancement via domain interaction modeling. However, many approaches overlook visual perceptual features and lack multimodal fusion with attention mechanisms, which can better focus on relevant signals from images, text, and other media, enriching user preference modeling across domains.

\subsection{LLM-based Recommendation System}
Large Language Models (LLMs) have gained significant traction in recommender systems due to their ability to leverage textual features of items and improve recommendation performance~\cite{geng2022recommendation, liu2023user,wu2023survey,qiu2024tfb,qiu2025duet,qiu2025tab}. Most existing LLM-based recommenders operate without fine-tuning, relying on pretrained knowledge to predict the next item~\cite{sun2023chatgpt, wang2023recmind}. For example, CHAT-REC~\cite{gao2023chat} uses ChatGPT to understand user preferences and deliver interactive, explainable recommendations. Similarly, GPT4Rec~\cite{li2023gpt4rec} employs GPT-2 to generate hypothetical "search queries" from user history, which are then processed by the BM25 search engine to retrieve relevant items. Another line of research focuses on fine-tuning LLMs for specific recommendation tasks. TALLRec~\cite{bao2023tallrec}, for instance, applies instruction-tuning to decide whether an item should be recommended, while BIGRec~\cite{bao2023bi} combines LLM-generated recommendations with collaborative filtering through an ensemble approach. Additionally, RecInterpreter~\cite{yang2023large} and LLaRA~\cite{liao2023llara} treat "sequential user behaviors" as a new modality for LLMs, aligning it with language representations. CoLLM~\cite{zhang2025collm} captures collaborative signals using an external model and integrates them into the LLM's token embedding space, enabling the use of collaborative embeddings.

While some studies like TALLRec~\cite{bao2023tallrec}, BIGRec~\cite{bao2023bi}, and LLM-Rec~\cite{tang2023one} acknowledge the cross-domain potential of LLMs, there remains a gap in frameworks explicitly designed for CDSR that combine LLM-based textual enrichment with multimodal fusion and domain-balanced attention.

Compared with recent representative methods, our work addresses several gaps. 
First, unlike LLMRec~\cite{wei2024llmrec}, which augments item text but does not explicitly mitigate domain imbalance, our framework applies a prompt-designed LLM augmentation strategy to generate domain-agnostic attributes that improve cross-domain semantic alignment. 
Second, in contrast to IFCDSR~\cite{wu2025image}, which fuses image and ID embeddings without enriched textual features, we unify visual, textual, and ID representations within a single multimodal fusion framework. 
Finally, whereas MAN~\cite{lin2024mixed} employs mixed attention without controlling domain contributions, our hierarchical attention mechanism explicitly balances the influence of each domain, preventing high-frequency domains from dominating recommendations.

\begin{figure}[t] 
\begin{center}
   \includegraphics[width=1.0\linewidth]{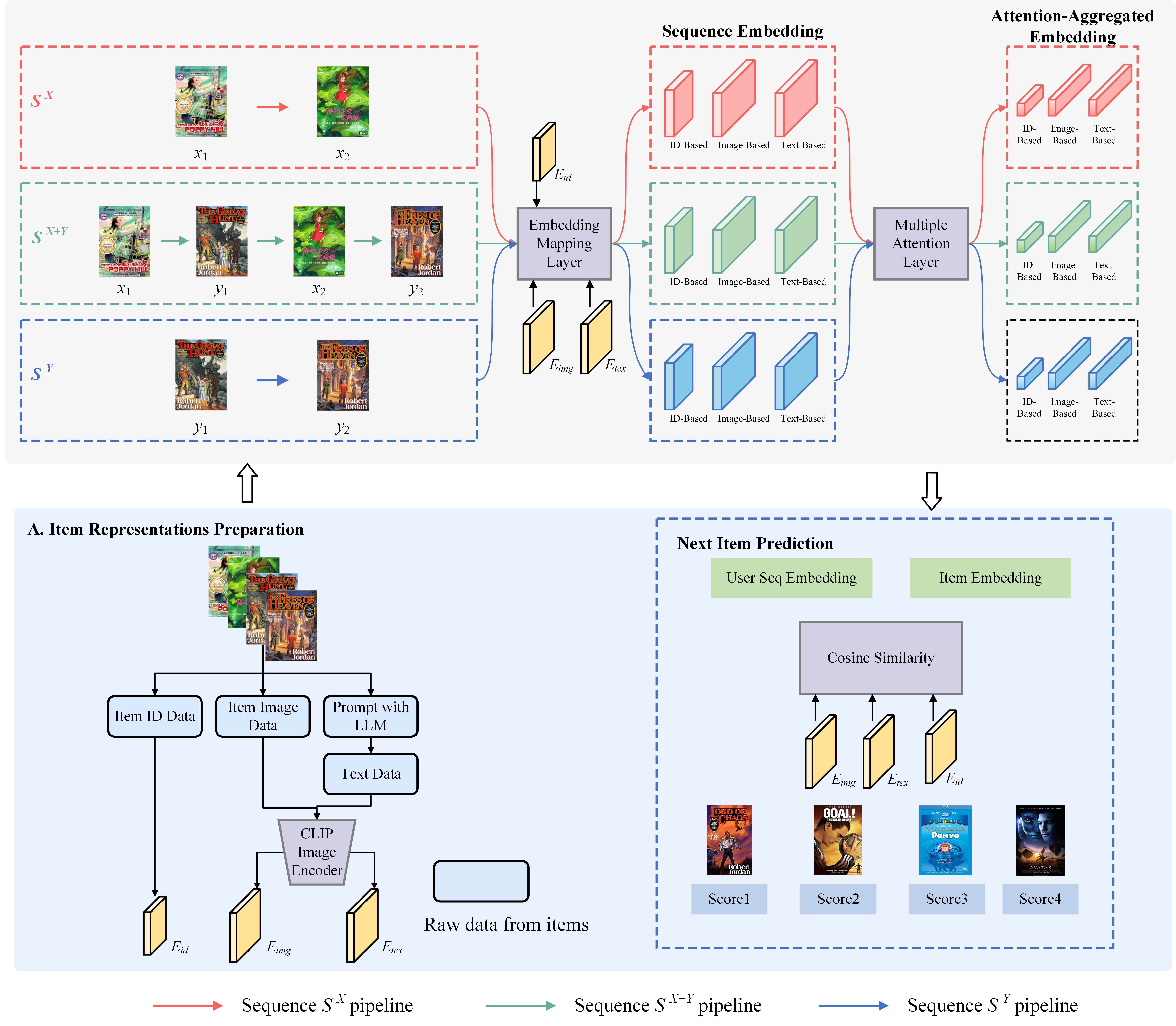}
\caption{
Overview of the proposed LLM-EMF framework. 
The Feature Preparation module generates ID-, image-, and text-based embeddings from domains $X$ and $Y$ using a learnable ID matrix, a frozen CLIP image encoder, and a text encoder. 
These embeddings are processed through multi-layer attention to model intra- and inter-sequence relationships, and cosine similarity with the embedding matrices is used for next-item prediction.
}
    \label{fig:framework}
\end{center}
\vspace{-0.4cm}
\end{figure}

\section{Methodology}
\label{sec:method}

\subsection{Problem Formulation}
In the CDSR task, user interaction sequences occur in domain $X$ and domain $Y$, with their item sets denoted as $\mathcal{X}$ and $\mathcal{Y}$, respectively. Let $\mathcal{S}$ represent the overall interaction sequence of the user in chronological order, which consists of three sub-sequences ${(S^X, S^Y, S^{X+Y}) \in \mathcal{S}}$. Specifically, let $S^X = \left[ x_1, x_2, \dots, x_{\left| S^X \right|} \right], x \in \mathcal{X}$ and $S^Y = \left[ y_1, y_2, \dots, y_{\left| S^Y \right|} \right], y \in \mathcal{Y}$ represent the interaction sequences within each domain, where $\left| \cdot \right|$ denotes the total number of items in each sequence.

Additionally, ${S^{X+Y} = {\left [ x_1, y_1, x_2,\dots , x_ {\left | S^X \right |}, \dots ,y_ {\left | S^Y \right |}  \right ]}}$ represents the merged sequence by combining $S^X$ and $S^Y$. In general, the goal of the CDSR task is to predict the probabilities of  candidate items across both domains and select the item with the highest probability as the next  recommendation.

\subsection{Overall framework}
As illustrated in Fig.~\ref{fig:framework}, the proposed LLM-EMF framework is designed to model cross-domain sequential preferences through a hierarchical architecture. The framework begins with item representation encoding, where items from domains $X$ and $Y$ are processed through parallel pathways to construct distinct representations. A learnable semantic embedding matrix $E_{id}$ is initialized to capture identity-specific features, while a frozen CLIP encoder generates perceptual embeddings $E_{img}$ and conceptual embeddings $E_{tex}$ from item images and textual descriptions, respectively. This multimodal encoding stage simulates the human capacity for integrating visual, semantic, and linguistic information.

The framework then proceeds to the preference simulation stage, where behavioral sequences $\mathcal{S}$ are processed. Each sequence $\mathcal{S}$ consists of three sub-sequences: $S^X$, $S^Y$, and $S^{X+Y}$, representing within-domain and cross-domain interactions. The embedding layer retrieves and restructures the multimodal embeddings ($E_{id}$, $E_{img}$, and $E_{tex}$) according to the sequence order, generating dynamic states for each sub-sequence. These states are further processed through hierarchical attention layers to model both intra-sequence and inter-sequence relationships, simulating the interplay between local working memory and global contextual integration.

Finally, the framework performs decision generation by comparing the sequence embeddings with the original multimodal embeddings ($E_{id}$, $E_{img}$, and $E_{tex}$) using cosine similarity. This process mimics human decision-making by combining evidence from multiple sensory modalities to predict the next item in the sequence.

\begin{figure*}[t] 
\begin{center}
   \includegraphics[width=1\linewidth]{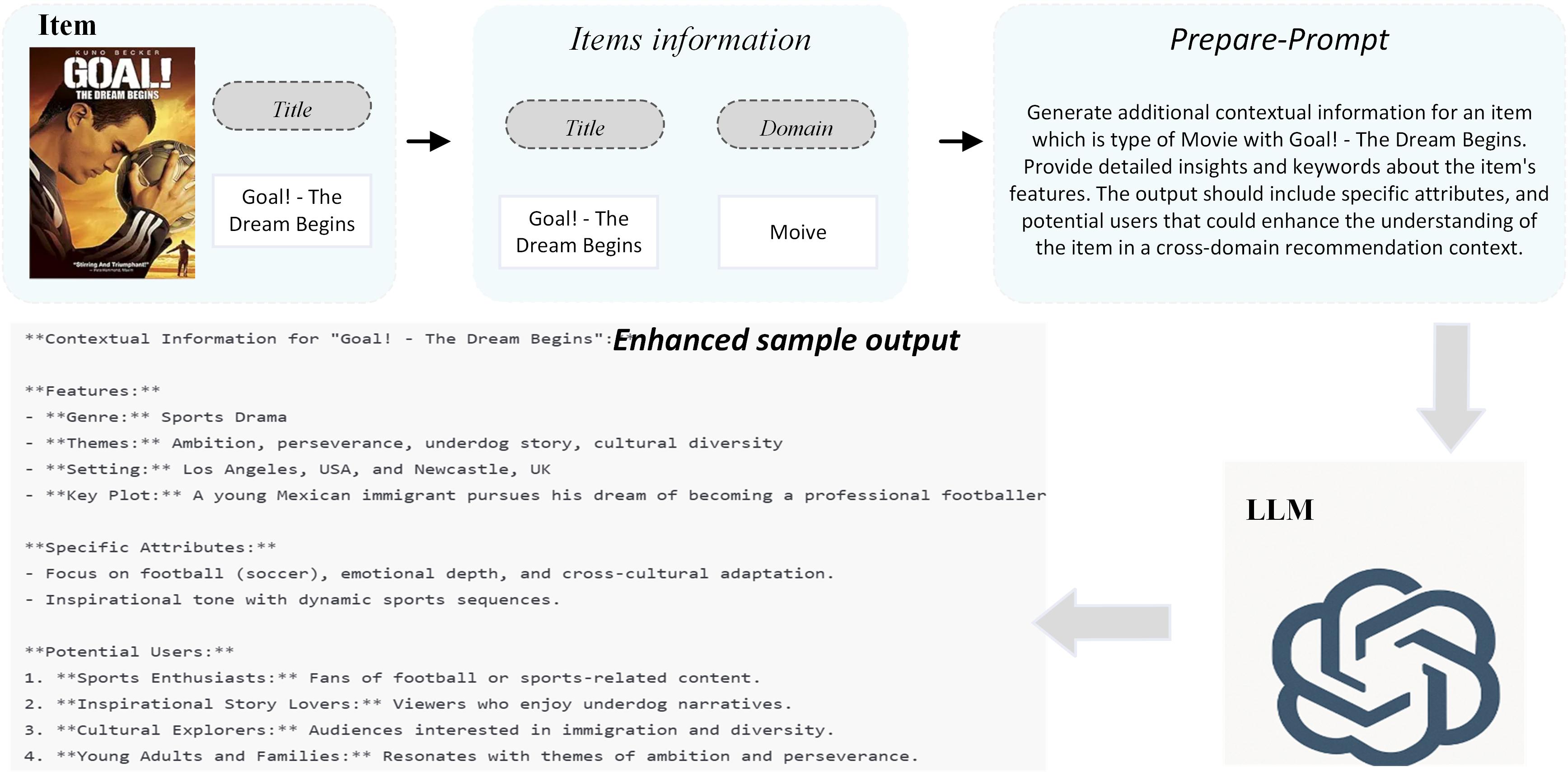}
   \caption{The prompt-and-generate pipeline of our method. In the showcase, the movie item is enhanced using the LLM with prompts to generate additional contextual information. As shown in the pipeline, the process begins by generating prompts, which are then input into the LLM. The output from the LLM consists of enhanced information, including key terms and a summary of the item, along with potential user interests. This enhanced information is subsequently used for the next step in text feature embedding.}
    \label{fig:prompt}
\end{center}
\end{figure*}
\subsection{Prompt with LLM}
In the CDSR setting, we denote user–item interaction sequences as $X_{item}$ and their corresponding item metadata (e.g., titles and domains) as $T_{info}$. As illustrated in Fig.~\ref{fig:prompt}, for each item in the dataset, a pre-defined template is used as a language command for Deepseek-r1~\cite{guo2025deepseek} to generate additional contextual knowledge. This approach ensures the relevance, diversity, and richness of item information, which is critical for enhancing item representations across multiple domains.

\begin{equation}
    P_{n} = \text{LLM}(commands),
\end{equation}

where $P_{n}$ represents the generated text prompt for enriching item representations, which is subsequently used as input for enhancing item embeddings. For each item, we design a unified template as the input command for Deepseek-r1~\cite{guo2025deepseek}, leveraging the item's title and description to generate additional contextual knowledge that further enriches the representation. This process requires only item-level metadata, making it both scalable and efficient.

For instance, the template is designed as follows:  
\textit{``Generate additional contextual information for an item of type [domain] with [title]. Provide detailed insights and keywords about the item's features. The output should include specific attributes and potential users that could enhance the understanding of the item in a cross-domain recommendation context.''}

This approach not only augments the item's metadata but also bridges gaps between domains by capturing domain-agnostic features and user preferences. By integrating the generated knowledge into the recommendation framework, we enable more comprehensive and accurate modeling of user behavior across diverse domains. The enriched embeddings, combined with our hierarchical attention mechanism, facilitate effective learning of both intra-domain and cross-domain relationships, ultimately improving the performance of the CDSR system.

\subsection{Visual and Textual Feature Integration}
\label{sec:Image}
We enhance user preferences by integrating multimodal features (image and textual) into the CDSR framework. This subsection provides a detailed explanation of feature preparation and sequence representation combined with image and textual features within a single domain, and applies this approach to domain $X$, domain $Y$, and domain $X+Y$.

\textit{1) Item Presentations Preparation with Image Feature:} 
First, we prepare the feature representations for all items. As shown on the left in Fig.~\ref{fig:framework}, we construct a learnable item matrix based on item IDs for all items in domain $X$ and $Y$, denoted as $E_{id} \in \mathbb{R}^{(|\mathcal{X}|+|\mathcal{Y}|) \times q}$. Here, $|\mathcal{X}|+|\mathcal{Y}|$ represents the total number of items from domains $X$ and $Y$, and $q$ is the learnable item embedding dimension. Simultaneously, we employ a pre-trained and frozen CLIP model to generate image embeddings for each item. These embeddings are used to form image matrix $E_{img} \in \mathbb{R}^{(|\mathcal{X}|+|\mathcal{Y}|) \times e}$, where $e$ is the image embedding dimension. These two matrices serve as the base for generating embeddings for sequences.

\textit{2) Item Presentations Preparation with Textual Feature:} 
To further enrich the item representations, we leverage textual information from item titles using the same frozen CLIP model. For each item, the CLIP text encoder generates textual embeddings based on the item's title, forming the textual matrix $E_{tex} \in \mathbb{R}^{(|\mathcal{X}|+|\mathcal{Y}|) \times e}$. This textual representation captures semantic information that complements the visual and ID-based features, enabling a more comprehensive understanding of user preferences.

\textit{3) Enhanced Similarity Score with Multimodal Features:}
We first generate a sequence representation to represent the user interaction sequence, which is then used to calculate the similarity with items for predicting the next item in the prediction process. As shown on the right in Fig.~\ref{fig:framework}, we propose an embedding layer to process the input sequences $\mathcal{S}$, which consists of three sub-sequences $S^X$, $S^Y$, and $S^{X+Y}$. These sequences include item IDs, images, and textual data. Accordingly, the embedding layer produces item ID-based, image-based, and textual-based embeddings for each input sequence. 

Once we obtain $E_{id}$, $E_{img}$, and $E_{tex}$, the item ID-based sequence embeddings $F_{id} \in \mathbb{R}^{|\mathcal{S}| \times q}$ for $\mathcal{S}$ can be generated by placing the appropriate embedding from $E_{id}$ in the order of the items in the sequence $\mathcal{S}$. Here, $|\mathcal{S}|$ denotes the total item number in $\mathcal{S}$. Similarly, the image-based sequence embeddings $F_{img} \in \mathbb{R}^{|\mathcal{S}| \times e}$ and textual-based sequence embeddings $F_{tex} \in \mathbb{R}^{|\mathcal{S}| \times e}$ are produced from $E_{img}$ and $E_{tex}$, respectively, based on the same sequence. 

For simplicity, we take the image-based sequence embeddings $F_{img}$ as an example to illustrate how LLM-EMF obtains the next item prediction. 
We employ an Attention Layer to obtain the enhanced sequence of image-based embeddings $H_{img} \in \mathbb{R}^{|\mathcal{S}| \times e}$, which captures the most relevant visual features from the input sequence for improved representation learning, as follows:
\begin{equation}
    H_{img} = Attention(F_{img}).
\end{equation}
Afterward, we extract the last embedding vector $h_{img} \in \mathbb{R}^{1 \times e}$ from $H_{img}$ as the sequence representation, which serves as the attention-aggregated embedding of the sequence. This vector captures the user preferences, specifically focusing on the most recent interaction within the sequence. Then, $h_{img}$ is compared against $E_{img}$ using cosine similarity. In this way, the alignment between user preferences and the embedding of each item across all domains is assessed. The similarity score is computed as follows:
\begin{equation} 
\begin{aligned}\label{eq:cos}
Sim(h_{img}, E_{img}) = \frac{h_{img} \cdot E_{img}^{T}}{\|h_{img}\|\|E_{img}^T\|},
\end{aligned}
\end{equation}
where $T$ denotes the transpose function. Here, a higher similarity score indicates that the sequence preference is more aligned with this item. In the same manner, we can obtain the item ID-based sequence representative vector $h_{id} \in \mathbb{R}^{1 \times q}$ and textual-based sequence representative vector $h_{tex} \in \mathbb{R}^{1 \times e}$, along with their respective similarity scores. In the next section, we will introduce how the multiple attention layer is used to fuse ID-based, image-based, and textual-based predictions for CDSR.  
\begin{figure}[t]
\centering
\includegraphics[width=1.0\linewidth]{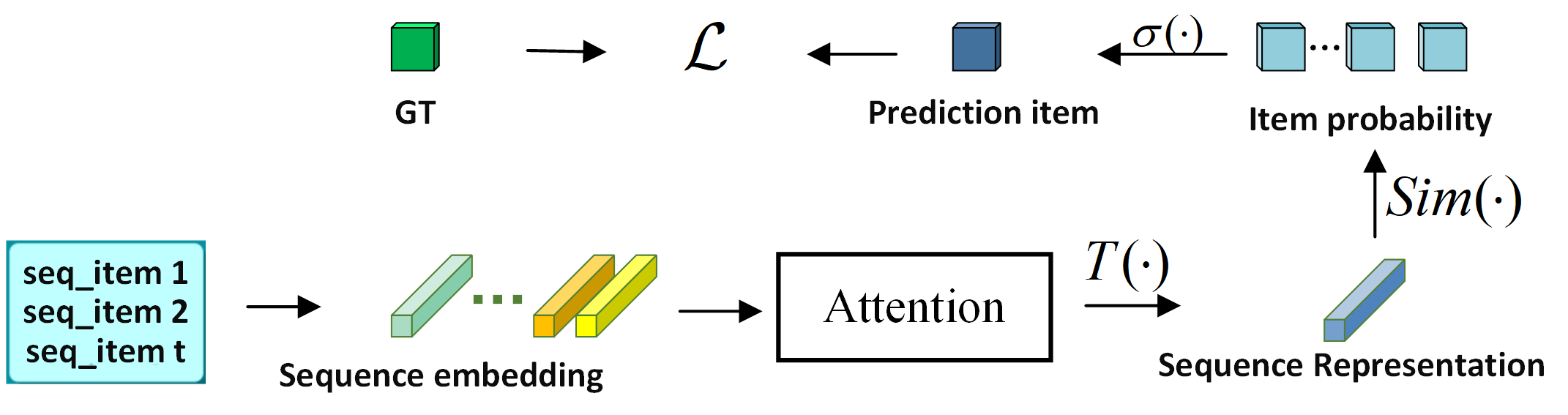}
\caption{The process of transforming a user sequence into a representation sequence. Initially, the user’s sequence of interactions is converted into an embedded representation. This sequence is then passed through an attention layer to capture both intra- and inter-sequence relationships. Finally, the attention-aggregated sequence representations are compared with item embeddings, and the item with the highest similarity score is selected as the predicted next item.}
\label{fig:detail}
\end{figure}
\subsection{Multiple Attention Mechanisms}
\label{sec:Attention}

In traditional sequential recommendation methods~\cite{gru4rec,sasrec}, behavioral sequences from domains $X$ and $Y$ are often processed indiscriminately, leading to potential dominance by the domain with higher behavioral frequency. This imbalance can skew the representation of user preferences. To address this issue, we propose a hierarchical attentional framework that separately processes sequences $S^X$, $S^Y$, and $S^{X+Y}$, ensuring balanced representation across domains. Our approach integrates item identity (ID), visual features, and textual embeddings to model user preferences more comprehensively. As illustrated in Fig.~\ref{fig:detail}, to model the interactions within each sequence more effectively, we incorporate a self-attention mechanism with query, key, and value (QKV) pairs. The attention mechanism allows the model to focus on relevant parts of the sequence when making predictions, capturing both local and global dependencies within the sequence.

The attention mechanism computes the output as a weighted sum of the values, where the weights are determined by the similarity between the query and the keys:
\begin{equation}
\text{Attention}(Q, K, V) = \text{softmax}\left(\frac{QK^T}{\sqrt{d_k}}\right) V,
\end{equation}
where $Q$, $K$, and $V$ are the query, key, and value matrices, respectively, and $d_k$ is the dimension of the key vectors. This mechanism is applied within each domain's sequence processing to capture both intra-domain and cross-domain relationships more effectively.

The framework generates nine sequence representations through multiple attention layers: 
\begin{equation}
h^{X}_{id}, h^{X}_{img}, h^{X}_{tex}, h^{Y}_{id}, h^{Y}_{img}, h^{Y}_{tex}, h^{X+Y}_{id}, h^{X+Y}_{img}, h^{X+Y}_{tex}. 
\end{equation}
These representations capture both intra-domain and cross-domain user preferences. For domain $X$, the prediction probability for the next item $x_{t+1}$ in sequence $S^X$ is computed as follows:
\begin{equation}
\mathrm{P}^X_{id}(x_{t+1} \mid \mathcal{S}) = \text{softmax}\left(\text{Sim}(h_{id}^X, E_{id}^X)\right), \quad x_t \in S^X,
\end{equation}
where $x_{t+1}$ denotes the predicted next item, $h_{id}^X$ is the ID-based sequence representation, and $E_{id}^X$ is the ID embedding matrix for domain $X$. Similarly, the prediction probabilities based on visual and textual features are calculated as:
\begin{equation}
\begin{aligned}
\mathrm{P}^X_{img}(x_{t+1} \mid \mathcal{S}) &= \text{softmax}\left(\text{Sim}(h_{img}^X, E_{img}^X)\right), \quad &x_t \in S^X,\\
\mathrm{P}^X_{tex}(x_{t+1} \mid \mathcal{S}) &= \text{softmax}\left(\text{Sim}(h_{tex}^X, E_{tex}^X)\right), \quad &x_t \in S^X,
\end{aligned}
\end{equation}
where $h_{img}^X$, $E_{img}^X$, $h_{tex}^X$, and $E_{tex}^X$ represent the image-based sequence representation, image embedding matrix, text-based sequence representation, and text embedding matrix, respectively. The final prediction probability for the next item combines ID, image, and text-based predictions:
\begin{align}
\mathrm{P}^X(x_{t+1} \mid \mathcal{S}) &= \alpha \mathrm{P}^X_{id}(x_{t+1} \mid \mathcal{S}) \nonumber \\
&\quad + \beta \mathrm{P}^X_{img}(x_{t+1} \mid \mathcal{S}) \nonumber \\
&\quad + (1-\alpha-\beta) \mathrm{P}^X_{tex}(x_{t+1} \mid \mathcal{S}),
\end{align}
where $\alpha$ and $\beta$ are weighting factors that balance the contributions of ID, visual, and textual features. This formulation captures structural, perceptual, and semantic aspects of user preferences.

To optimize the framework, we minimize the following loss function for domain $X$:
\begin{equation}
\small
\mathcal{L}^X = \sum_{x_t \in S^X} -\log \mathrm{P}^X(x_{t+1} \mid \mathcal{S}).
\label{sec_math_eq:single}
\end{equation}
The same method is applied to domains $Y$ and $X+Y$, yielding loss terms $\mathcal{L}^Y$ and $\mathcal{L}^{X+Y}$. The final loss is a weighted combination of these terms:
\begin{equation}
\mathcal{L} = \mathcal{L}^{X} + \lambda_{1}\mathcal{L}^{Y} + \lambda_{2}\mathcal{L}^{X+Y},
\end{equation}
where $\lambda_1$ and $\lambda_2$ control the relative importance of each domain.

During evaluation, we aggregate predictions from all domains to recommend the next item. Let $X$ be the target domain and $Y$ the source domain. The combined prediction score for item $x_i$ is computed as:
\begin{equation}
\mathrm{P}(x_i|S) = \mathcal{P}^{X}(x_i|S) + \lambda_{1} \mathcal{P}^{Y}(x_i|S) + \lambda_{2} \mathcal{P}^{X+Y}(x_i|S).
\end{equation}
The recommended item is selected by maximizing the prediction score within the target domain $X$:
\begin{equation}
\mathrm{argmax}_{x_i \in \mathcal{X}} \mathrm{P}(x_i|S).
\end{equation}
This process allows the model to incorporate cross-domain information effectively, leveraging the predictions from both the target domain and the source domain. By weighting the contributions from each domain, the model can refine its recommendation, taking into account the richer information that might come from the source domain while focusing on the target domain for the final prediction.

The weighting factors $\lambda_1$ and $\lambda_2$ allow the model to control how much influence the source domain ($Y$) and the combined domain ($X+Y$) have on the recommendation for the target domain. This flexibility in adjusting the weights is crucial for handling different types of datasets or domain-specific characteristics, ensuring that the model adapts to varying degrees of relevance between domains.

Finally, by selecting the item with the highest prediction score from the target domain, the model aims to provide the most contextually appropriate recommendation, which is critical for real-world recommendation systems that need to balance both accuracy and relevance across multiple domains.

\begin{table}[t]
\centering
\caption{Statistics of two CDSR scenarios.}
\vspace{0.4cm}
\begin{tabular}{cccccc}
\toprule
\textbf{Scenarios}  &\textbf{\#Items} &\textbf{\#Train} &\textbf{\#Valid}  &\textbf{\#Test} &\textbf{Avg.length}\\ \midrule
Food  &29,207  &\multirow{2}{*}{34,117} &2,722   &2,747   &\multirow{2}{*}{9.91}     \\ 
Kitchen  &34,886 &\multirow{2}{*}{}  &5,451   &5,659  &\multirow{2}{*}{}     \\
\midrule
Movie  &36,845  &\multirow{2}{*}{58,515} &2,032  &1,978  &\multirow{2}{*}{11.98}    \\ 
Book  &63,937 &\multirow{2}{*}{}  &5,612  &5,730 &\multirow{2}{*}{}   \\ 
\midrule

\end{tabular}
\label{sec_exp_tab:dataset}
\end{table}

\begin{table*}[t]
\footnotesize
\centering
\caption{Experimental results (\%) on the Food-Kitchen scenario.}
\vspace{0.3cm}
\label{tab:foodkitchen}
\setlength\tabcolsep{4.5pt} 
\resizebox{\columnwidth}{!}{ 
\begin{tabular}{lcccccc}
\toprule
\multirow{2}{*}{Model (Food-Kitchen)} &
\multirow{2}{*}{MRR} &\multicolumn{2}{c}{NDCG} & \multirow{2}{*}{MRR} &\multicolumn{2}{c}{NDCG} \\
\cmidrule(r){3-4}\cmidrule(l){6-7} &(Food)& @5 & @10  &(Kitchen)& @5  & @10   \\
\midrule
NCF-MLP~\cite{he2017neural} &  4.49  &   3.94   &  4.51 &  2.18 &  1.57  &  2.03  \\
$\pi$-Net~\cite{pinet}   &   7.68  &   7.32  & 8.13 &
 3.53 &   2.98  &  3.73  \\

GRU4Rec~\cite{gru4rec}   &   5.79  &   5.48   &  6.13 &
 3.06 &  2.55  &  3.10  \\
SASRec~\cite{sasrec}   &   7.30  &   6.90   &  7.79  &
 3.79 &  3.35  &  3.93  \\ 
SR-GNN~\cite{srgnn}   &    7.84  &   7.58  & 8.35   &
  4.01 &   3.47   &  4.13  \\
PSJNet~\cite{PSJnet}   &   8.33  &   8.07  & 8.77 &
4.10 &  3.68  &  4.32  \\
MIFN~\cite{mifn}  & 8.55  &   8.28  & 9.01   &
  4.09 &   3.57   &  4.29  \\
Tri-CDR~\cite{ma2024triple} &  8.35 &  8.18 &   8.89  &  
4.29  &  3.63 &   4.33  \\
MAN~\cite{lin2024mixed}              & 8.65 & 8.42 & 9.41 & 4.78 & 4.47 & 5.05 \\
LLMRec~\cite{wei2024llmrec}        & 9.05 & 8.86 & 10.03 & 5.02 & 4.73 & 5.34 \\
IFCDSR~\cite{wu2025image} &  9.05 &  8.85 &   9.92  &  
4.95  &  4.56 &   5.24  \\
\textbf{Ours LLM-EMF} &\textbf{9.24} &\textbf{9.05} &\textbf{10.21}
&\textbf{5.13} &\textbf{4.84} &\textbf{5.43}\\
\bottomrule
\end{tabular}
}
\end{table*}

\begin{table*}[t]
\footnotesize
\centering
\caption{Experimental results (\%) on the Movie-Book scenario.}
\vspace{0.3cm}
\label{tab:moviebook}
\setlength\tabcolsep{4.5pt} 
\resizebox{\columnwidth}{!}{ 
\begin{tabular}{lcccccc}
\toprule
\multirow{2}{*}{Model (Movie-Book)} &
\multirow{2}{*}{MRR} &\multicolumn{2}{c}{NDCG} & \multirow{2}{*}{MRR} &\multicolumn{2}{c}{NDCG} \\
\cmidrule(r){3-4}\cmidrule(l){6-7} &(Movie)  & @5 & @10  &(Book)& @5  & @10   \\
\midrule
NCF-MLP~\cite{he2017neural} &  3.05 &   2.26    &   2.96 & 1.43   &   1.06 &  1.26  \\
$\pi$-Net~\cite{pinet}   &   4.16  &   3.72  & 4.17 &
2.17 &   1.84   &  2.03  \\
GRU4Rec~\cite{gru4rec}   &  3.83 &   3.14 &  3.73  &
 1.68 &  1.34   &  1.52  \\

SASRec~\cite{sasrec}   &   3.79 &   3.23 &  3.69  &
 1.81 &  1.41   &  1.71  \\

SR-GNN~\cite{srgnn}   &  3.85 &  3.27   &  3.78 &
  1.78 &  1.40   &  1.66  \\

PSJNet~\cite{PSJnet}  &  4.63 &  4.06 &   4.76 & 
 2.44  &  2.07 &   2.35  \\

MIFN~\cite{mifn}  &  5.05 &  4.21 &   5.20  &  
2.51  &  2.12 &   2.31  \\
MAN~\cite{lin2024mixed}              & 5.95 & 5.06 & 5.45 & 2.64 & 2.34 & 2.57 \\
Tri-CDR~\cite{ma2024triple} &  5.15 &  4.62 &   5.05  &  
2.32  &  2.08 &   2.22  \\
LLMRec~\cite{wei2024llmrec}        & 6.21 & 5.29 & 5.71 & 2.80 & 2.48 & 2.74 \\
IFCDSR~\cite{wu2025image}            & 6.08 & 5.02 & 5.86 & 2.75 & 2.37 & 2.65 \\

\textbf{Ours LLM-EMF} &\textbf{6.32}  &\textbf{5.35} &\textbf{5.78}  
&\textbf{2.86} &\textbf{2.53} &\textbf{2.79} \\
\bottomrule
\end{tabular}

}
\end{table*}
\vspace{0.4cm}

\section{Experiments}
\label{sec:Experiments}

In this section, we introduce the experimental setup, providing detailed descriptions of the dataset, evaluation metrics, and implementation specifics. We then compare our method with state-of-the-art approaches on the Amazon dataset~\cite{wei2021contrastive}. Finally, we conduct ablation studies to validate the effectiveness of the proposed method.

\textbf{Dataset and Evaluated Metric.} Our experiments are conducted on the Amazon dataset~\cite{wei2021contrastive}, which serves as the foundation for constructing cross-domain sequential recommendation (CDSR) scenarios. The dataset offers a rich collection of user-item interactions across a variety of domains, making it ideal for modeling cross-domain behaviors. Building on previous works~\cite{pinet, mifn}, we select four distinct domains to create two CDSR scenarios for our experiments: ``Food-Kitchen'' and ``Movie-Book.'' These domains were chosen to reflect real-world applications where users frequently engage with multiple categories, allowing us to investigate the efficacy of cross-domain knowledge transfer in sequential recommendation settings. The process of preparing the data begins by extracting users who have interacted with items in both domains. To ensure meaningful analysis, we filter out users and items with fewer than 10 interactions, focusing only on those with sufficient behavioral data. This filtering step is crucial for ensuring the robustness of our results, as sparse data can lead to unreliable recommendations. Furthermore, in order to meet the sequential constraints necessary for our study, we retain cross-domain interaction sequences that contain at least three items from each domain. This ensures that each sequence provides enough context for learning temporal dependencies and domain transitions. For the training/validation/test split, we adopt a temporal strategy to preserve the sequential nature of the data. Specifically, we divide the latest interaction sequences into validation and test sets, while the remaining sequences are allocated to the training set. This approach helps simulate a real-world scenario, where past user behavior informs future recommendations. The statistics of our refined datasets for the CDSR scenarios, including the number of users, items, and interactions in each domain, are summarized in Tab.~\ref{sec_exp_tab:dataset}.

To evaluate the performance of our model, we adopt two widely used evaluation metrics in recommendation systems: the Mean Reciprocal Rank (MRR) metric~\cite{mrr}, which measures the rank of the first relevant item, and the Normalized Discounted Cumulative Gain (NDCG\@5, 10) metric ~\cite{ndcg}.
MRR is a metric that assesses the ranking quality by calculating the average of the reciprocal ranks of the first relevant item in each user's recommendation list. NDCG, on the other hand, takes into account both the position and relevance of the recommended items, making it particularly suitable for evaluating ranked lists. These metrics are standard in the evaluation of recommendation systems, providing a comprehensive measure of recommendation quality. The results obtained from these metrics allow us to compare our model's performance across different CDSR scenarios and demonstrate its effectiveness in capturing cross-domain interactions.

\begin{figure}[t]
\centering
\includegraphics[width=1.0\linewidth]{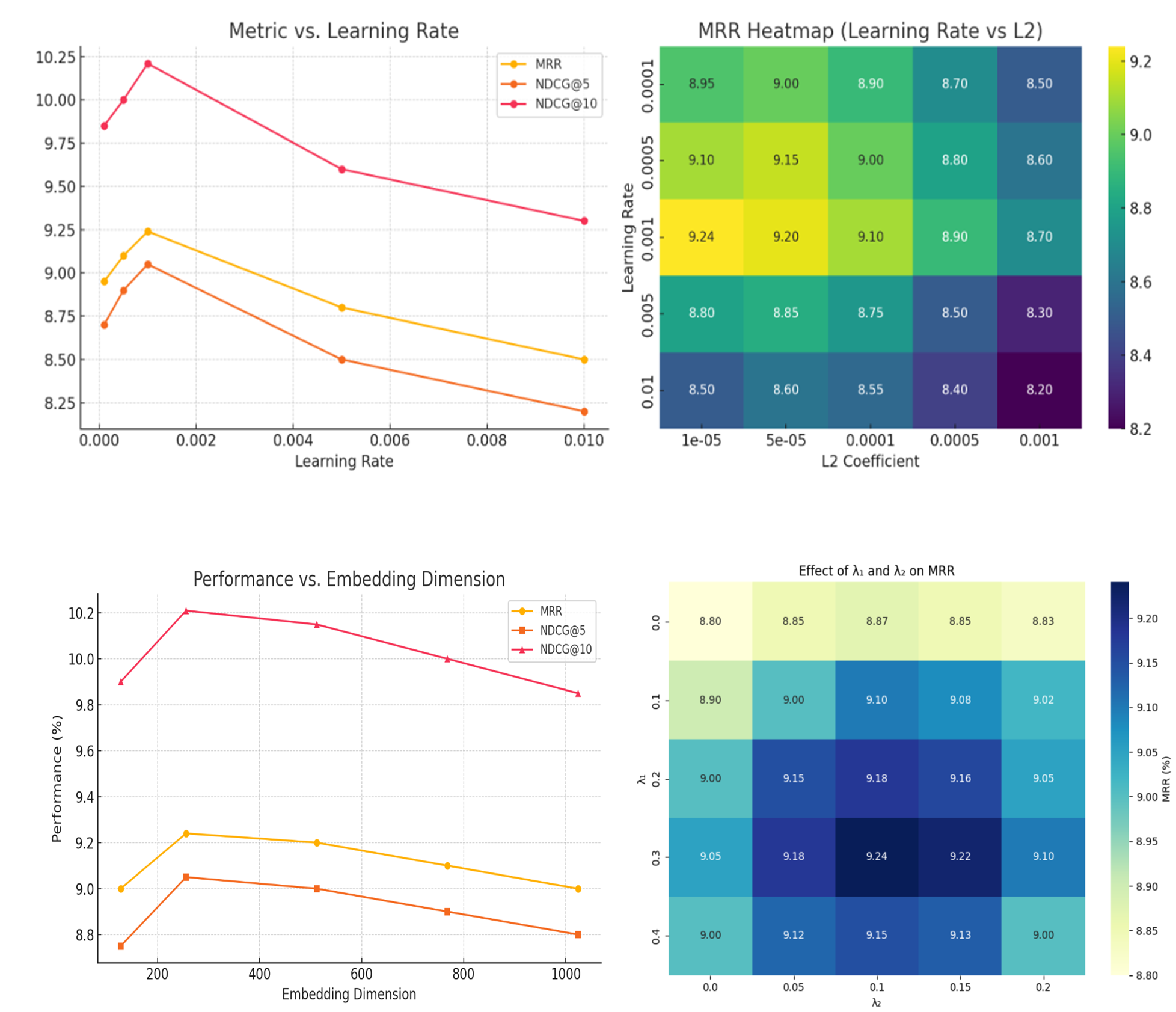}
\vspace{0.3cm} 
\caption{Impact of Hyperparameters on Final Performance on Movie Data}
\label{fig:hyper}
\end{figure}

\textbf{Implementation Details.} For fair comparisons, we adopt the same hyperparameter settings as in previous works~\cite{mifn}. We conduct experiments with various hyperparameter settings, as shown in Figure~\ref{fig:hyper}. Specifically, we set the learnable embedding size $q$ to 256, the CLIP-based image and text embedding size $e$ to 512, and the mini-batch size to 256, with a dropout rate of 0.3. The values of $\lambda_{1}$ and $\lambda_{2}$ are set to 0.3 and 0.1, respectively. The $L_2$ regularization coefficient is selected from $\{0.00001, 0.00005, 0.0001, 0.0005, 0.001\}$, and the learning rate is chosen from $\{0.0001, 0.0005, 0.001, 0.005, 0.01\}$. Training is conducted for 100 epochs using an NVIDIA 4090 GPU, with the Adam optimizer~\cite{adam} to update all parameters. To ensure robustness and reproducibility, we perform a grid search over the hyperparameter space, evaluating the model's performance on a validation set at each step. The selected hyperparameters are those that yield the best validation performance. Additionally, we employ early stopping with a patience of 10 epochs to prevent overfitting, monitoring the validation loss during training. This approach ensures that the model generalizes well to unseen data while maintaining computational efficiency. The use of the Adam optimizer~\cite{adam} is particularly advantageous due to its adaptive learning rate mechanism, which helps stabilize training and accelerate convergence. Furthermore, the NVIDIA 4090 GPU provides the necessary computational power to handle the large-scale datasets and complex model architectures efficiently. These settings collectively ensure that our experiments are conducted under optimal conditions, enabling meaningful comparisons with state-of-the-art methods in the field.

\subsection{Performance Comparisons}

We evaluate our proposed LLM-EMF method against a comprehensive set of state-of-the-art (SOTA) approaches on the ``Food-Kitchen'' and ``Movie-Book'' CDSR scenarios. The experimental results are summarized in Tab.~\ref{tab:foodkitchen} and Tab.~\ref{tab:moviebook}, respectively. Our approach consistently outperforms all existing methods across both scenarios, demonstrating its effectiveness in handling cross-domain recommendation tasks. In the Food-Kitchen scenario, our LLM-EMF model achieves the highest scores across all evaluation metrics. Specifically, it attains an MRR of 9.24\% for Food and 5.13\% for Kitchen, significantly outperforming the closest competitor, MIFN, which achieves MRR scores of 8.55\% and 4.09\%, respectively. In terms of NDCG@5 and NDCG@10, LLM-EMF also leads with 9.05\% and 10.21\% for Food, and 4.84\% and 5.43\% for Kitchen. The performance of traditional methods such as ItemKNN, NCF-MLP, and BPRMF is notably lower, highlighting the limitations of these approaches in capturing complex user-item interactions in cross-domain settings. More advanced models like $\pi$-Net, GRU4Rec, and SASRec show improved performance but still fall short of LLM-EMF. This suggests that our model's ability to leverage large language models (LLMs) for enhanced feature extraction and cross-domain knowledge transfer is a key factor in its superior performance. In the Movie-Book scenario, LLM-EMF again demonstrates its superiority, achieving MRR scores of 6.32\% for Movie and 2.86\% for Book. These results are significantly higher than those of the next best model, MIFN, which achieves MRR scores of 5.05\% and 2.51\%, respectively. Similarly, in terms of NDCG@5 and NDCG@10, LLM-EMF leads with 5.35\% and 5.78\% for Movie, and 2.53\% and 2.79\% for Book.

\begin{table}[t]
\centering
\caption{Ablation study on the movie dataset}
\vspace{0.4cm}
\label{tab:ablation}
\begin{adjustbox}{width=1.0\linewidth}
\begin{tabular}{ccccccc}
\toprule
 original-framework & Textual Fusion &LLM-enhanced & Visual Fusion &  MRR \\
\midrule
\checkmark & & &  & 5.03 \\
  \checkmark & \checkmark&  & &  5.88 \\
  \checkmark & \checkmark&\checkmark  & &  6.01 \\
  \checkmark &  &  & \checkmark & 6.08 \\
  \checkmark & \checkmark&  & \checkmark&  6.15 \\
 \checkmark & \checkmark& \checkmark & \checkmark & 6.32 \\
\bottomrule
\end{tabular}
\end{adjustbox}
\end{table}

\subsection{Ablation Studies}
We conducted an ablation study to systematically evaluate the contributions of our proposed components: \textit{Textual Fusion}, \textit{LLM-enhanced}, and \textit{Visual Fusion}. As illustrated in Tab.~\ref{tab:ablation}, we first established a baseline using the original framework without any additional modules, achieving an MRR of 5.03. The introduction of \textit{Textual Fusion} alone resulted in a performance gain of 0.85 points, increasing the MRR to 5.88, which highlights the importance of leveraging textual information for cross-domain recommendation tasks. When the \textit{LLM-enhanced} module was added alongside \textit{Textual Fusion}, the MRR further improved to 6.01, demonstrating the effectiveness of leveraging large language models to enrich textual representations. Notably, the inclusion of \textit{Visual Fusion} alone significantly boosted the MRR to 6.08, underscoring the critical role of visual signals in enhancing the model's understanding of user preferences. The combination of \textit{Textual Fusion} and \textit{Visual Fusion} achieved an MRR of 6.15, indicating that textual and visual information complement each other in capturing cross-domain user behavior. Finally, the full framework, incorporating \textit{Textual Fusion}, \textit{LLM-enhanced}, and \textit{Visual Fusion}, achieved the highest MRR of 6.32. This result demonstrates the cumulative impact of integrating multiple modalities and advanced language models for improving recommendation performance.

\section{Conclusion}

In this study, we propose a novel framework, named LLM-EMF, for cross-domain sequential recommendation. Unlike traditional methods that focus solely on temporal patterns within user interaction sequences, our approach leverages LLM knowledge to enrich textual information while also utilizing the powerful visual and textual understanding capabilities of the CLIP model to integrate multimodal representations. Specifically, we design prompts to combine the title and domain information, which are then used to generate augmented text. Both the images and augmented text are processed through CLIP to generate embeddings, which are subsequently fused with a learnable item matrix to enhance item representations. By combining these innovations, our LLM-EMF framework achieves state-of-the-art performance in CDSR, demonstrating the effectiveness of integrating multimodal signals and LLM knowledge in sequential recommendation tasks.

\section{Acknowledgements}
This work was supported by the National Natural Science Foundation of China (No. 62471405, 62331003, 62301451), Suzhou Basic Research Program (SYG202316) and XJTLU REF-22-01-010, XJTLU AI University Research Centre, Jiangsu Province Engineering Research Centre of Data Science and Cognitive Computation at XJTLU and SIP AI innovation platform (YZCXPT2022103).
\bibliography{mybib}

@String(JCST  = {J. Comput. Sci. Tech.})

@String(NIPS= {Adv. Neural Inf. Process. Syst.})

@String(ICLR = {Int. Conf. Learn. Represent.})

@String(IJCAI = {Int. Joint Conf. Artificial Intell.})

@String(AAAI = {AAAI Conf. Artif. Intell.})

@String(NIPS={Int. Conf. Neur. Info. Process. Sys.})

@article{PSJnet,
author = {Sun, Wenchao and Ren, Pengjie and Lin, Yujie and Ma, Muyang and Chen, Zhumin and Ren, Zhaochun and Ma, Jun and de Rijke, Maarten},
journal = {IEEE Transactions on Knowledge and Data Engineering (TKDE)},
title = {Parallel Split-Join Networks for Shared Account Cross-domain Sequential Recommendations},
year = {2022}
}

@article{Zhuangict,
  title={Sequential Recommendation via Cross-Domain Novelty Seeking Trait Mining},
  author={Fuzhen Zhuang and Yingmin Zhou and Haochao Ying and Fuzheng Zhang and Xiang Ao and Xing Xie and Qing He and Hui Xiong},
  journal={Journal of Computer Science and Technology (JCST)},
  year={2020},
}

@inproceedings{liu2024fedbcgd,
  title={Fedbcgd: Communication-efficient accelerated block coordinate gradient descent for federated learning},
  author={Liu, Junkang and Shang, Fanhua and Liu, Yuanyuan and Liu, Hongying and Li, Yuangang and Gong, YunXiang},
  booktitle={Proceedings of the 32nd ACM International Conference on Multimedia},
  pages={2955--2963},
  year={2024}
}

@article{xiao2025cross,
  title={Cross-domain recommendation via interest-aware pseudo-overlapping user alignment},
  author={Xiao, Shitong and Chen, Rui and Song, Hongtao and Han, Qilong},
  journal={Expert Systems with Applications},
  volume={292},
  pages={128638},
  year={2025},
  publisher={Elsevier}
}

@article{liang2025graphical,
  title={Graphical contrastive learning for multi-interest sequential recommendation},
  author={Liang, Shunpan and Kong, Qianjin and Lei, Yu and Li, Chen},
  journal={Expert Systems with Applications},
  volume={259},
  pages={125285},
  year={2025},
  publisher={Elsevier}
}

@article{wu2025impact,
  title={Impact of alleviating misinformation: an impulsive buying-aware model for sequential recommendation},
  author={Wu, Hongchen and Fang, Xiaochang and Li, Hongxuan and Sun, Jie and Jing, Jing and Zhang, Lin and Meng, Yihong and Jing, Zhaorong and Zhang, Huaxiang},
  journal={Expert Systems with Applications},
  pages={128513},
  year={2025},
  publisher={Elsevier}
}

@inproceedings{wei2024llmrec,
  title={Llmrec: Large language models with graph augmentation for recommendation},
  author={Wei, Wei and Ren, Xubin and Tang, Jiabin and Wang, Qinyong and Su, Lixin and Cheng, Suqi and Wang, Junfeng and Yin, Dawei and Huang, Chao},
  booktitle={Proceedings of the 17th ACM international conference on web search and data mining},
  pages={806--815},
  year={2024}
}

@inproceedings{liuimproving,
  title={Improving Generalization in Federated Learning with Highly Heterogeneous Data via Momentum-Based Stochastic Controlled Weight Averaging},
  author={Liu, Junkang and Liu, Yuanyuan and Shang, Fanhua and Liu, Hongying and Liu, Jin and Feng, Wei},
  booktitle={Forty-second International Conference on Machine Learning},
  year={2025}
}

@article{ndcg,
  title={Cumulated gain-based evaluation of IR techniques},
  author={Kalervo J{\"a}rvelin and Jaana Kek{\"a}l{\"a}inen},
  journal={ACM Transactions on Information Systems (TOIS)},
  year={2002},
}

@article{markovsr,
  title={An MDP-Based Recommender System},
  author={Guy Shani and David Heckerman and Ronen I. Brafman},
  journal={Journal of Machine Learning Research (JMLR)},
  year={2002}
}

@article{mifn,
  title={Mixed Information Flow for Cross-domain Sequential Recommendations},
  author={Muyang Ma and Pengjie Ren and Zhumin Chen and Zhaochun Ren and Lifan Zhao and Jun Ma and M. de Rijke},
  journal={ACM Transactions on Knowledge Discovery from Data (TKDD)},
  year={2022},
}

@article{cnn,
  title={Convolutional networks for images, speech, and time series},
  author={LeCun, Yann and Bengio, Yoshua and others},
  journal={The handbook of brain theory and neural networks},
  year={1995}
}

@inproceedings{rendle2010factorizing,
  title={Factorizing personalized markov chains for next-basket recommendation},
  author={Rendle, Steffen and Freudenthaler, Christoph and Schmidt-Thieme, Lars},
  booktitle={WWW},
  pages={811--820},
  year={2010}
}

@inproceedings{mrr,
  title={The TREC-8 Question Answering Track Report},
  author={Ellen M. Voorhees},
  booktitle={Text REtrieval Conference (TREC)},
  year={1999}
}

@inproceedings{dagcn,
  title={DA-GCN: A Domain-aware Attentive Graph Convolution Network for Shared-account Cross-domain Sequential Recommendation},
  author={Lei Guo and Li Tang and Tong Chen and Lei Zhu and Quoc Viet Hung Nguyen and Hongzhi Yin},
  booktitle={International Joint Conference on Artificial Intelligence (IJCAI)},
  year={2021}
}

@inproceedings{srgnn,
  title={Session-based recommendation with graph neural networks},
  author={Wu, Shu and Tang, Yuyuan and Zhu, Yanqiao and Wang, Liang and Xie, Xing and Tan, Tieniu},
  booktitle={AAAI Conference on Artificial Intelligence (AAAI)},
  year={2019}
}

@inproceedings{caser,
  title={Personalized Top-N Sequential Recommendation via Convolutional Sequence Embedding},
  author={Jiaxi Tang and Ke Wang},
  booktitle={ACM International Conference on Web Search and Data Mining (WSDM)},
  year={2018}
}

@inproceedings{kang2018self,
	title={Self-attentive sequential recommendation},
	author={Kang, Wang-Cheng and McAuley, Julian},
	booktitle={2018 IEEE International Conference on Data Mining (ICDM)},
	pages={197--206},
	year={2018},
	organization={IEEE}
}

@inproceedings{recguru,
  title={RecGURU: Adversarial Learning of Generalized User Representations for Cross-Domain Recommendation},
  author={Li, Chenglin and Zhao, Mingjun and Zhang, Huanming and Yu, Chenyun and Cheng, Lei and Shu, Guoqiang and Kong, Beibei and Niu, Di},
  booktitle={ACM International Conference on Web Search and Data Mining (WSDM)},
  year={2022}
}

@inproceedings{conet,
  title={Conet: Collaborative Cross Networks for Cross-Domain Recommendation},
  author={Hu, Guangneng and Zhang, Yu and Yang, Qiang},
  booktitle={ACM International Conference on Information and Knowledge Management (CIKM)},
  year={2018}
}

@inproceedings{narm,
  title={Neural Attentive Session-based Recommendation},
  author={J. Li and Pengjie Ren and Zhumin Chen and Zhaochun Ren and Tao Lian and Jun Ma},
  booktitle={ACM International Conference on Information and Knowledge Management (CIKM)},
  year={2017}
}

@inproceedings{DIN,
  title={Deep interest network for click-through rate prediction},
  author={Zhou, Guorui and Zhu, Xiaoqiang and Song, Chenru and Fan, Ying and Zhu, Han and Ma, Xiao and Yan, Yanghui and Jin, Junqi and Li, Han and Gai, Kun},
  booktitle={KDD},
  pages={1059--1068},
  year={2018}
}

@inproceedings{SURGE,
  title={Sequential Recommendation with Graph Neural Networks},
  author={Chang, Jianxin and Gao, Chen and Zheng, Yu and Hui, Yiqun and Niu, Yanan and Song, Yang and Jin, Depeng and Li, Yong},
  booktitle={Proceedings of the 44th International ACM SIGIR Conference on Research and Development in Information Retrieval},
  pages={378--387},
  year={2021}
}

@inproceedings{lin2023dual,
  title={Dual-interest Factorization-heads Attention for Sequential Recommendation},
  author={Lin, Guanyu and Gao, Chen and Zheng, Yu and Chang, Jianxin and Niu, Yanan and Song, Yang and Li, Zhiheng and Jin, Depeng and Li, Yong},
  booktitle={Proceedings of the ACM Web Conference 2023},
  pages={917--927},
  year={2023}
}

@inproceedings{wu2025image,
  title={Image fusion for cross-domain sequential recommendation},
  author={Wu, Wangyu and Song, Siqi and Qiu, Xianglin and Huang, Xiaowei and Ma, Fei and Xiao, Jimin},
  booktitle={Companion Proceedings of the ACM on Web Conference 2025},
  pages={2196--2202},
  year={2025}
}

@inproceedings{lin2024mixed,
  title={Mixed attention network for cross-domain sequential recommendation},
  author={Lin, Guanyu and Gao, Chen and Zheng, Yu and Chang, Jianxin and Niu, Yanan and Song, Yang and Gai, Kun and Li, Zhiheng and Jin, Depeng and Li, Yong and others},
  booktitle={Proceedings of the 17th ACM international conference on web search and data mining},
  pages={405--413},
  year={2024}
}

@inproceedings{lin2022dual,
  title={Dual contrastive network for sequential recommendation},
  author={Lin, Guanyu and Gao, Chen and Li, Yinfeng and Zheng, Yu and Li, Zhiheng and Jin, Depeng and Li, Yong},
  booktitle={Proceedings of the 45th international ACM SIGIR conference on research and development in information retrieval},
  pages={2686--2691},
  year={2022}
}

@inproceedings{adam,
  title={Adam: A Method for Stochastic Optimization},
  author={Kingma, P. Diederik and Ba, Lei Jimmy},
  booktitle={International Conference on Learning Representations (ICLR)},
  year={2015}
}

@inproceedings{gru4rec,
  title={Session-based Recommendations with Recurrent Neural Networks},
  author={Bal{\'a}zs Hidasi and Alexandros Karatzoglou and Linas Baltrunas and Domonkos Tikk},
  booktitle={International Conference on Learning Representations (ICLR)},
  year={2015},
}

@inproceedings{cao2022contrastive,
  title={Contrastive cross-domain sequential recommendation},
  author={Cao, Jiangxia and Cong, Xin and Sheng, Jiawei and Liu, Tingwen and Wang, Bin},
  booktitle={Proceedings of the 31st ACM International Conference on Information \& Knowledge Management},
  pages={138--147},
  year={2022}
}

@article{ma2024triple,
  title={Triple sequence learning for cross-domain recommendation},
  author={Ma, Haokai and Xie, Ruobing and Meng, Lei and Chen, Xin and Zhang, Xu and Lin, Leyu and Zhou, Jie},
  journal={ACM Transactions on Information Systems},
  volume={42},
  number={4},
  pages={1--29},
  year={2024},
  publisher={ACM New York, NY}
}

@inproceedings{kddsemi,
  title={SEMI: A Sequential Multi-Modal Information Transfer Network for E-Commerce Micro-Video Recommendations},
  author={Lei, Chenyi and Liu, Yong and Zhang, Lingzi and Wang, Guoxin and Tang, Haihong and Li, Houqiang and Miao, Chunyan},
  booktitle={ACM Knowledge Discovery and Data Mining (KDD)},
  year={2021}
}

@article{transmarkov,
  title={Translation-based Recommendation},
  author={Ruining He and Wang-Cheng Kang and Julian McAuley},
  booktitle={ACM Conference on Recommender Systems (RecSys)},
  year={2017}
}

@inproceedings{sasrec,
  title={Self-Attentive Sequential Recommendation},
  author={Wang-Cheng Kang and Julian McAuley},
  booktitle={IEEE International Conference on Data Mining (ICDM)},
  year={2018},
}

@inproceedings{markov,
  title={Fusing Similarity Models with Markov Chains for Sparse Sequential Recommendation},
  author={Ruining He and Julian McAuley},
  booktitle={IEEE International Conference on Data Mining (ICDM)},
  year={2016},
}

@incollection{CDR,
	title = {Cross-domain recommender systems},
	pages = {919--959},
	booktitle = {Recommender systems handbook},
	publisher = {Springer},
	author = {Cantador, Iván and Fernández-Tobías, Ignacio and Berkovsky, Shlomo and Cremonesi, Paolo},
	year = {2015},
}

@inproceedings{singh_relational_2008,
	title = {Relational learning via collective matrix factorization},
	pages = {650--658},
	booktitle = {Proceedings of the 14th {ACM} {SIGKDD} international conference on Knowledge discovery and data mining},
	author = {Singh, Ajit P. and Gordon, Geoffrey J.},
	year = {2008},
}

@article{transfer,
	title = {A survey on transfer learning},
	volume = {22},
	pages = {1345--1359},
	number = {10},
	journaltitle = {{IEEE} Transactions on knowledge and data engineering},
	author = {Pan, Sinno Jialin and Yang, Qiang},
	year = {2009},
	note = {Publisher: {IEEE}},
}

@inproceedings{MiNet,
  title={MiNet: Mixed Interest Network for Cross-Domain Click-Through Rate Prediction},
  author={Ouyang, Wentao and Zhang, Xiuwu and Zhao, Lei and Luo, Jinmei and Zhang, Yu and Zou, Heng and Liu, Zhaojie and Du, Yanlong},
  booktitle={Proceedings of the 29th ACM International Conference on Information \& Knowledge Management},
  pages={2669--2676},
  year={2020}
}

@inproceedings{itemCST,
	title = {Transfer learning in collaborative filtering for sparsity reduction},
	volume = {24},
	booktitle = {Proceedings of the {AAAI} Conference on Artificial Intelligence},
	author = {Pan, Weike and Xiang, Evan and Liu, Nathan and Yang, Qiang},
	year = {2010},
	note = {Issue: 1},
}

@inproceedings{long_dual_2012,
	title = {Dual transfer learning},
	pages = {540--551},
	booktitle = {Proceedings of the 2012 {SIAM} International Conference on Data Mining},
	publisher = {{SIAM}},
	author = {Long, Mingsheng and Wang, Jianmin and Ding, Guiguang and Cheng, Wei and Zhang, Xiang and Wang, Wei},
	year = {2012},
}

@article{he_dual_2016,
	title = {Dual learning for machine translation},
	volume = {29},
	pages = {820--828},
	journaltitle = {Advances in neural information processing systems},
	author = {He, Di and Xia, Yingce and Qin, Tao and Wang, Liwei and Yu, Nenghai and Liu, Tie-Yan and Ma, Wei-Ying},
	year = {2016},
}

@inproceedings{pinet,
	title = {$\pi$-Net: A parallel information-sharing network for shared-account cross-domain sequential recommendations},
	shorttitle = {π-Net},
	pages = {685--694},
	booktitle = {Proceedings of the 42nd International {ACM} {SIGIR} Conference on Research and Development in Information Retrieval},
	author = {Ma, Muyang and Ren, Pengjie and Lin, Yujie and Chen, Zhumin and Ma, Jun and Rijke, Maarten de},
	year = {2019}
}

@inproceedings{wei2021contrastive,
  title={Contrastive learning for cold-start recommendation},
  author={Wei, Yinwei and Wang, Xiang and Li, Qi and Nie, Liqiang and Li, Yan and Li, Xuanping and Chua, Tat-Seng},
  booktitle={Proceedings of the 29th ACM International Conference on Multimedia},
  pages={5382--5390},
  year={2021}
}

@article{SRs,
	title = {Sequential recommender systems: challenges, progress and prospects},
	shorttitle = {Sequential recommender systems},
	journaltitle = {{arXiv} preprint {arXiv}:2001.04830},
	author = {Wang, Shoujin and Hu, Liang and Wang, Yan and Cao, Longbing and Sheng, Quan Z. and Orgun, Mehmet},
	year = {2019},
}

@inproceedings{GRU,
  title={Empirical evaluation of gated recurrent neural networks on sequence modeling},
  author={Chung, Junyoung and Gulcehre, Caglar and Cho, Kyunghyun and Bengio, Yoshua},
  booktitle={NIPS 2014 Workshop on Deep Learning, December 2014},
  year={2014}
}

@article{LSTM,
  title={Long short-term memory},
  author={Hochreiter, Sepp and Schmidhuber, J{\"u}rgen},
  journal={Neural computation},
  volume={9},
  number={8},
  pages={1735--1780},
  year={1997},
  publisher={MIT Press}
}

@inproceedings{vaswani2017attention,
  title={Attention is all you need},
  author={Vaswani, Ashish and Shazeer, Noam and Parmar, Niki and Uszkoreit, Jakob and Jones, Llion and Gomez, Aidan N and Kaiser, {\L}ukasz and Polosukhin, Illia},
  booktitle = {NeurIPS},
  pages={5998--6008},
  year={2017}
}

@inproceedings{DIEN,
  title={Deep interest evolution network for click-through rate prediction},
  author={Zhou, Guorui and Mou, Na and Fan, Ying and Pi, Qi and Bian, Weijie and Zhou, Chang and Zhu, Xiaoqiang and Gai, Kun},
  booktitle={AAAI},
  pages={5941--5948},
  year={2019}
}

@inproceedings{geng2022recommendation,
  title={Recommendation as language processing (rlp): A unified pretrain, personalized prompt \& predict paradigm (p5)},
  author={Geng, Shijie and Liu, Shuchang and Fu, Zuohui and Ge, Yingqiang and Zhang, Yongfeng},
  booktitle={Proceedings of the 16th ACM conference on recommender systems},
  pages={299--315},
  year={2022}
}

@inproceedings{liu2023user,
  title={User Behavior Modeling with Deep Learning for Recommendation: Recent Advances},
  author={Liu, Weiwen and Guo, Wei and Liu, Yong and Tang, Ruiming and Wang, Hao},
  booktitle={Proceedings of the 17th ACM Conference on Recommender Systems},
  pages={1286--1287},
  year={2023}
}

@article{wu2023survey,
  title={A Survey on Large Language Models for Recommendation},
  author={Wu, Likang and Zheng, Zhi and Qiu, Zhaopeng and Wang, Hao and Gu, Hongchao and Shen, Tingjia and Qin, Chuan and Zhu, Chen and Zhu, Hengshu and Liu, Qi and others},
  journal={arXiv preprint arXiv:2305.19860},
  year={2023}
}

@article{sun2023chatgpt,
  title={Is ChatGPT good at search? investigating large language models as re-ranking agents},
  author={Sun, Weiwei and Yan, Lingyong and Ma, Xinyu and Wang, Shuaiqiang and Ren, Pengjie and Chen, Zhumin and Yin, Dawei and Ren, Zhaochun},
  journal={arXiv preprint arXiv:2304.09542},
  year={2023}
}

@article{wang2023recmind,
  title={Recmind: Large language model powered agent for recommendation},
  author={Wang, Yancheng and Jiang, Ziyan and Chen, Zheng and Yang, Fan and Zhou, Yingxue and Cho, Eunah and Fan, Xing and Huang, Xiaojiang and Lu, Yanbin and Yang, Yingzhen},
  journal={arXiv preprint arXiv:2308.14296},
  year={2023}
}

@inproceedings{cai2024relation,
  title={Relation-Fused Attention in Knowledge Graphs For Recommendation},
  author={Cai, Huanyi and Wu, Wangyu and Chai, Bosong and Zhang, Yafeng},
  booktitle={International Conference on Neural Information Processing},
  pages={285--299},
  year={2024},
  organization={Springer}
}

@article{gao2023chat,
  title={Chat-rec: Towards interactive and explainable llms-augmented recommender system},
  author={Gao, Yunfan and Sheng, Tao and Xiang, Youlin and Xiong, Yun and Wang, Haofen and Zhang, Jiawei},
  journal={arXiv preprint arXiv:2303.14524},
  year={2023}
}

@article{li2023gpt4rec,
  title={GPT4Rec: A generative framework for personalized recommendation and user interests interpretation},
  author={Li, Jinming and Zhang, Wentao and Wang, Tian and Xiong, Guanglei and Lu, Alan and Medioni, Gerard},
  journal={arXiv preprint arXiv:2304.03879},
  year={2023}
}

@inproceedings{bao2023tallrec,
  title={Tallrec: An effective and efficient tuning framework to align large language model with recommendation},
  author={Bao, Keqin and Zhang, Jizhi and Zhang, Yang and Wang, Wenjie and Feng, Fuli and He, Xiangnan},
  booktitle={Proceedings of the 17th ACM Conference on Recommender Systems},
  pages={1007--1014},
  year={2023}
}

@article{bao2023bi,
  title={A bi-step grounding paradigm for large language models in recommendation systems},
  author={Bao, Keqin and Zhang, Jizhi and Wang, Wenjie and Zhang, Yang and Yang, Zhengyi and Luo, Yancheng and Chen, Chong and Feng, Fuli and Tian, Qi},
  journal={ACM Transactions on Recommender Systems},
  year={2023},
  publisher={ACM New York, NY}
}

@article{yang2023large,
  title={Large language model can interpret latent space of sequential recommender},
  author={Yang, Zhengyi and Wu, Jiancan and Luo, Yanchen and Zhang, Jizhi and Yuan, Yancheng and Zhang, An and Wang, Xiang and He, Xiangnan},
  journal={arXiv preprint arXiv:2310.20487},
  year={2023}
}

@inproceedings{qiu2024tfb,
title   = {TFB: Towards Comprehensive and Fair Benchmarking of Time Series Forecasting Methods},
author  = {Xiangfei Qiu and Jilin Hu and Lekui Zhou and Xingjian Wu and Junyang Du and Buang Zhang and Chenjuan Guo and Aoying Zhou and Christian S. Jensen and Zhenli Sheng and Bin Yang},
booktitle = {Proc. {VLDB} Endow.},
pages   = {2363--2377},
year    = {2024}
}

@article{li2024towards,
  title={Towards Visual-Prompt Temporal Answer Grounding in Instructional Video},
  author={Li, Shutao and Li, Bin and Sun, Bin and Weng, Yixuan},
  journal={IEEE transactions on pattern analysis and machine intelligence},
  volume={46},
  number={12},
  pages={8836--8853},
  year={2024}
}

@article{li2024distinct,
  title={Distinct but correct: generating diversified and entity-revised medical response},
  author={Li, Bin and Sun, Bin and Li, Shutao and Chen, Encheng and Liu, Hongru and Weng, Yixuan and Bai, Yongping and Hu, Meiling},
  journal={Science China Information Sciences},
  volume={67},
  number={3},
  pages={132106},
  year={2024},
  publisher={Springer}
}

@article{li2024towards2,
  title={Towards better Chinese-centric neural machine translation for low-resource languages},
  author={Li, Bin and Weng, Yixuan and Xia, Fei and Deng, Hanjun},
  journal={Computer Speech \& Language},
  volume={84},
  pages={101566},
  year={2024},
  publisher={Elsevier}
}

@article{hou2024invdiff,
  title={InvDiff: Invariant Guidance for Bias Mitigation in Diffusion Models},
  author={Hou, Min and Wu, Yueying and Xu, Chang and Huang, Yu-Hao and Bai, Chenxi and Wu, Le and Bian, Jiang},
  journal={arXiv preprint arXiv:2412.08480},
  year={2024}
}

@article{li2025bridge,
  title={Bridge: Bootstrapping text to control time-series generation via multi-agent iterative optimization and diffusion modelling},
  author={Li, Hao and Huang, Yuhao and Xu, Chang and Schlegel, Viktor and Jiang, Renhe and Batista-Navarro, Riza and Nenadic, Goran and Bian, Jiang},
  journal={arXiv preprint arXiv:2503.02445},
  year={2025}
}

@inproceedings{qiu2025duet,
title   = {DUET: Dual Clustering Enhanced Multivariate Time Series Forecasting},
author  = {Xiangfei Qiu and Xingjian Wu and Yan Lin and Chenjuan Guo and Jilin Hu and Bin Yang},
booktitle = {SIGKDD},
pages     = {1185-1196},
year    = {2025}
}

@inproceedings{qiu2025tab,
title   = {TAB: Unified Benchmarking of Time Series Anomaly Detection Methods},
author  = {Xiangfei Qiu and Zhe Li and Wanghui Qiu and Shiyan Hu and Lekui Zhou and Xingjian Wu and Zhengyu Li and Chenjuan Guo and Aoying Zhou and Zhenli Sheng and Jilin Hu and Christian S. Jensen and Bin Yang},
booktitle = {Proc. {VLDB} Endow.},
year    = {2025}
}

@article{liao2023llara,
  title={Llara: Aligning large language models with sequential recommenders},
  author={Liao, Jiayi and Li, Sihang and Yang, Zhengyi and Wu, Jiancan and Yuan, Yancheng and Wang, Xiang},
  journal={CoRR},
  year={2023}
}

@article{zhang2025collm,
  title={Collm: Integrating collaborative embeddings into large language models for recommendation},
  author={Zhang, Yang and Feng, Fuli and Zhang, Jizhi and Bao, Keqin and Wang, Qifan and He, Xiangnan},
  journal={IEEE Transactions on Knowledge and Data Engineering},
  year={2025},
  publisher={IEEE}
}

@article{tang2023one,
  title={One model for all: Large language models are domain-agnostic recommendation systems},
  author={Tang, Zuoli and Huan, Zhaoxin and Li, Zihao and Zhang, Xiaolu and Hu, Jun and Fu, Chilin and Zhou, Jun and Li, Chenliang},
  journal={arXiv preprint arXiv:2310.14304},
  year={2023}
}

@inproceedings{he2017neural,
  title={Neural collaborative filtering},
  author={He, Xiangnan and Liao, Lizi and Zhang, Hanwang and Nie, Liqiang and Hu, Xia and Chua, Tat-Seng},
  booktitle={Proceedings of the 26th international conference on world wide web},
  pages={173--182},
  year={2017}
}

@article{guo2025deepseek,
  title={Deepseek-r1: Incentivizing reasoning capability in llms via reinforcement learning},
  author={Guo, Daya and Yang, Dejian and Zhang, Haowei and Song, Junxiao and Zhang, Ruoyu and Xu, Runxin and Zhu, Qihao and Ma, Shirong and Wang, Peiyi and Bi, Xiao and others},
  journal={arXiv preprint arXiv:2501.12948},
  year={2025}
}

@inproceedings{jiang2023structure,
  title={Structure-aware surface reconstruction via primitive assembly},
  author={Jiang, Jingen and Zhao, Mingyang and Xin, Shiqing and Yang, Yanchao and Wang, Hanxiao and Jia, Xiaohong and Yan, Dong-Ming},
  booktitle={Proceedings of the IEEE/CVF International Conference on Computer Vision},
  pages={14171--14180},
  year={2023}
}

\end{document}